\documentclass[conference]{IEEEtran}
\usepackage{cite}
\usepackage{amsmath,amssymb,amsfonts}
\usepackage{algorithmic}
\usepackage{graphicx}
\usepackage{textcomp}
\usepackage{xcolor}
\def\BibTeX{{\rm B\kern-.05em{\sc i\kern-.025em b}\kern-.08em
    T\kern-.1667em\lower.7ex\hbox{E}\kern-.125emX}}

\ifCLASSOPTIONcompsoc
    \usepackage[caption=false, font=normalsize, labelfont=sf, textfont=sf]{subfig}
\else
    \usepackage[caption=false, font=footnotesize]{subfig}
\fi

\usepackage{mathtools}
\DeclarePairedDelimiter\ceil{\lceil}{\rceil}

\begin{document}

\title{kMatrix: A Space Efficient Streaming Graph Summarization Technique}

\author{\IEEEauthorblockN{Oshan Mudannayake}
    \IEEEauthorblockA{\textit{University of Colombo School of Computing}\\
        Colombo, Sri Lanka \\
        oshan.ivantha@gmail.com}
    \and
    \IEEEauthorblockN{Nalin Ranasinghe}
    \IEEEauthorblockA{\textit{University of Colombo School of Computing}\\
        Colombo, Sri Lanka \\
        dnr@ucsc.cmb.ac.lk}
}

\maketitle

\begin{abstract}
    The amount of collected information on data repositories has vastly increased with the advent of the internet. It has become increasingly complex to deal with these massive data streams due to their sheer volume and the throughput of incoming data. Many of these data streams are mapped into graphs, which helps discover some of their properties. However, due to the difficulty in processing massive streaming graphs, they are summarized such that their properties can be approximately evaluated using the summaries. gSketch, TCM, and gMatrix are some of the major streaming graph summarization techniques. Our primary contribution is devising kMatrix, which is much more memory efficient than existing streaming graph summarization techniques. We achieved this by partitioning the allocated memory using a sample of the original graph stream. Through the experiments, we show that kMatrix can achieve a significantly less error for the queries using the same space as that of TCM and gMatrix.  
\end{abstract}


\section{Introduction}

Massive-scale datasets are becoming increasingly common today. The growth of the number of users who are actively using digital devices connected to the internet has vastly affected this phenomenon. Also, there lies an interest in researchers to solve the problems which involve large datasets. Most of these datasets could be mapped into graphs to extract useful information, giving rise to the need for processing massive scale graphs. There are many practical scenarios where massive scale graphs are applied such as social networks, network traffic data, and road networks.

It is much easier to work with graphs when they are static and small. However, most of the natural graphs that are being encountered in the real world are dynamic. It becomes increasingly complex to handle the graph as the velocity with which its edges get updated increases. Large scale dynamic natural graphs are used by many companies today. Google uses the PageRank algorithm\cite{brin_anatomy_1998, page_pagerank_nodate} to map the links between the web pages. Facebook has a massive graph with trillions of edges\cite{ching_one_2015}, depicting the interactions of each user on the platform.

With the size of the massive scale graphs, it is difficult to evaluate their properties even after partitioning into multiple nodes. The graphs have to be summarized so that important information regarding the underlying dataset can be inferred easily.

Being applied in a wide range of industrial and research applications, realtime property evaluation of streaming and dynamic natural graphs is a critical requirement in many scenarios. Graph summarization plays a significant role in this as it reduces the computational resources required to evaluate the properties in a rather massive scale streaming graph. It would be beneficial for many sectors if the process of summarizing streaming graphs were made efficient.

In this work, we propose an improved streaming graph summarization technique; kMatrix. It can outperform the existing state of the art summarization sketches by efficiently using the available memory to answer the queries more accurately. We also show that kMatrix is generally faster than the other sketches in handling the graph streams.
\section{Background}

Graphs can be divided into static graphs and streaming (dynamic) graphs. Static graphs do not change while streaming graphs are the ones that get updated over a time interval. These update operations could be insertion or deletion of nodes and edges.

Determining the properties of streaming graphs is a relatively strenuous task than static graphs as they are continually evolving. Thus the traditional graph algorithms cannot be run on streaming graphs due to their dynamic nature. Getting a static snapshot of a streaming graph at a specific timestamp and running conventional graph algorithms on it is one way of addressing this issue. However, this may not be a suitable remedy as the time taken to process the graph’s massive snapshot can render the result less valuable in a time-critical scenario. This process is made even more difficult with the speed with which the graph is being updated. High throughput of update queries requires any other types of queries to be run efficiently and as fast as possible in an unblocking manner. Therefore if there is a need for processing the graph while streaming, separate streaming graph algorithms have to be devised\cite{mcgregor_graph_2014}. 

Since real-world streaming graphs could grow very large in size, they are often stored as partitions in different machines over a network rather than in a single location. It is difficult to evaluate the properties of a graph with high volume and throughput even after the partitioning process, as the whole graph would have to be processed despite the partitioning. Graph summarization is a technique used in dealing with these massive graphs taking the limitations mentioned above into account. 

In graph summarization, we reduce the complexity of a graph while retaining only some of its properties. These summaries often incur an error when queried due to the loss of information. When the same algorithm is executed on a graph summary and its original graph, the two results are expected to be approximately equal. Here, the error depends on the compression ratio and various other factors. This tradeoff in accuracy is usually worth it for real-world graphs such as social networks when considering the computational cost incurred in obtaining exact answers. Most of the time, the cost of obtaining an exact solution is so high that it is impossible to do so even if the need arises. 

Summarizing a graph can have many benefits\cite{liu_graph_2018} apart from the speedup of graph algorithms and queries, such as, reduction of data volume and storage\cite{seo_effective_2018}, visualization\cite{dunne_motif_2013, jin_eco_nodate}, noise elimination\cite{zhang_discovery-driven_2010}, privacy preservation\cite{shoaran_zero-knowledge_2013}. 

Graph summarization has a wide range of industrial and research applications. Some of them are clustering\cite{cilibrasi_clustering_2005}, classification\cite{hutchison_compression_2006}, community detection\cite{chakrabarti_fully_nodate}, outlier detection\cite{smets_odd_2011, akoglu_opavion_2012}, pattern set mining\cite{mampaey_tell_2011} and finding sources of infection in large graphs\cite{prakash_spotting_2012}. Throughout this work, our aim lies in query optimization through graph summarization.

Streaming graph summarization is much more complex than summarizing a static graph due to the constant data flow. Since the underlying graph is updated continuously, the summarization process also has to be done in realtime. Almost any static graph summarization technique can be used with a streaming graph snapshot in a specific timestamp. However, mining information using aggregate time snapshots of data could prove to be a less than ideal solution when considering massive data streams. Thus sophisticated sparsification techniques have to be derived in order to summarize streaming graphs.
\section{Related Work}

\subsection{CountMin}

\begin{figure}[htbp]
    \centerline{\includegraphics[width=0.5\textwidth]{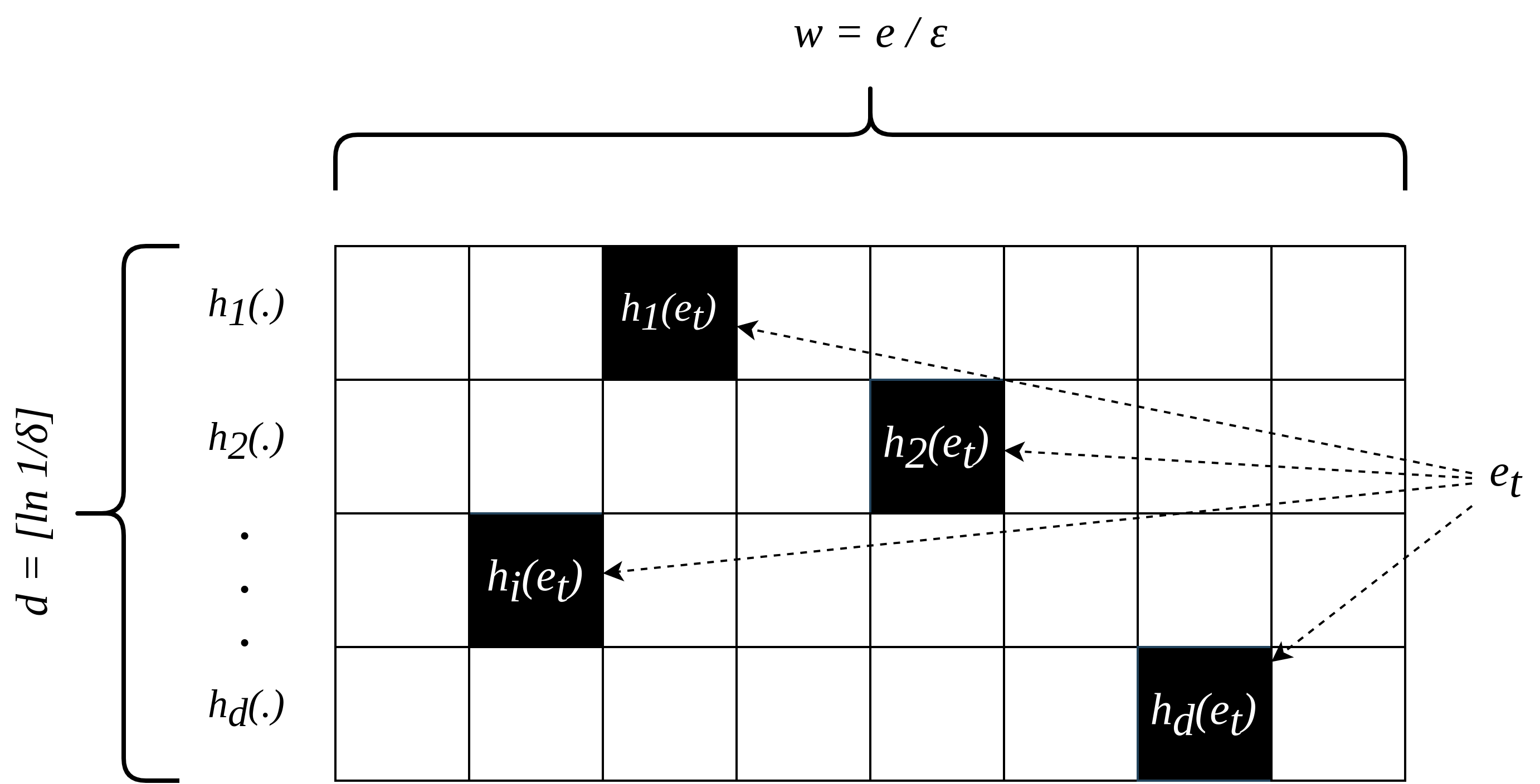}}
    \caption{CountMin sketch\cite{zhao_gsketch:_2011}}
    \label{fig:countmin}
\end{figure}

CountMin\cite{cormode_improved_2003} is a 2-dimensional data structure that is used for frequency approximation queries. It has a width of \(w = \ceil{e / \epsilon}\) and a depth of \(d = \ceil{ln(1 / \delta)}\). Here the \(e\) is the base of the natural logarithm while \(\epsilon\) and \(\delta\) are user-specified constants. The underlying idea is to hash the aggregated frequencies of the edges using multiple hash functions into predefined blocks, as indicated in Fig.~\ref{fig:countmin}. Any incoming edge \(e_t\) at timestamp \(t\) will get hashed into each row using its hash function \(h_d\). A CountMin sketch will have a fixed memory allocation of \(w \cdot d\) throughout its lifespan. Irrespective of the volume of the data stored in the sketch, the initial memory allocation will not change. Thus the accuracy of the queries will decrease as more and more data is inserted into the sketch. Despite the weaknesses, CountMin can be considered as a good generalized summarization sketch as many other current techniques are geared towards specific graph computation scenarios. However, the CountMin approach is not restricted to streaming graphs but other applications as well\cite{cormode_improved_2003}. 

\subsection{gSketch}

gSketch\cite{zhao_gsketch:_2011} is an extension of CountMin data structure. But unlike the CountMin sketch, this is specifically geared towards summarizing graph streams. gSketch is based on one of the below two assumptions.

\begin{itemize}
    \item A sample of the graph stream is available.
    \item Samples of both the graph stream and the query workload are available.
\end{itemize}

\begin{figure}[htbp]
    \centerline{\includegraphics[width=0.5\textwidth]{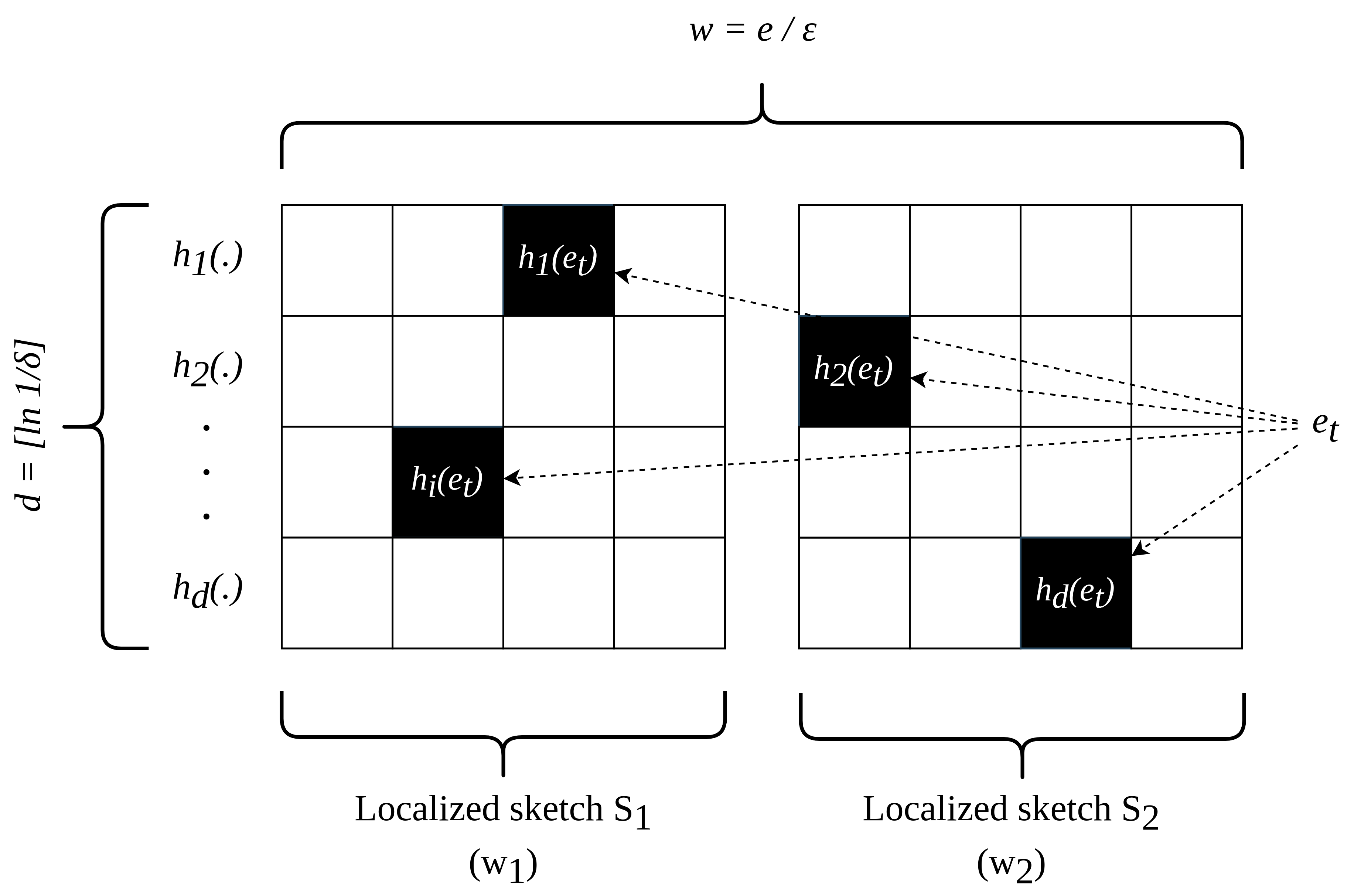}}
    \caption{gSketch sketch}
    \label{fig:gsketch}
\end{figure}

In CountMin, one global sketch is created for the entire stream. By doing so, it fails to take advantage of any structural properties present in the graph stream. gSketch tries to avoid this by considering the underlying structure of the graph stream using a sample. It then proceeds to partition its allocated space, as indicated in Fig.~\ref{fig:gsketch}. The goal of this partitioning step aims to maintain a sufficient frequency uniformity within each localized sketch in a way such that the combined error of the quarry estimations over the entire graph is kept at a minimum.

\subsection{TCM}

\begin{figure}[htbp]
    \centerline{\includegraphics[width=0.35\textwidth]{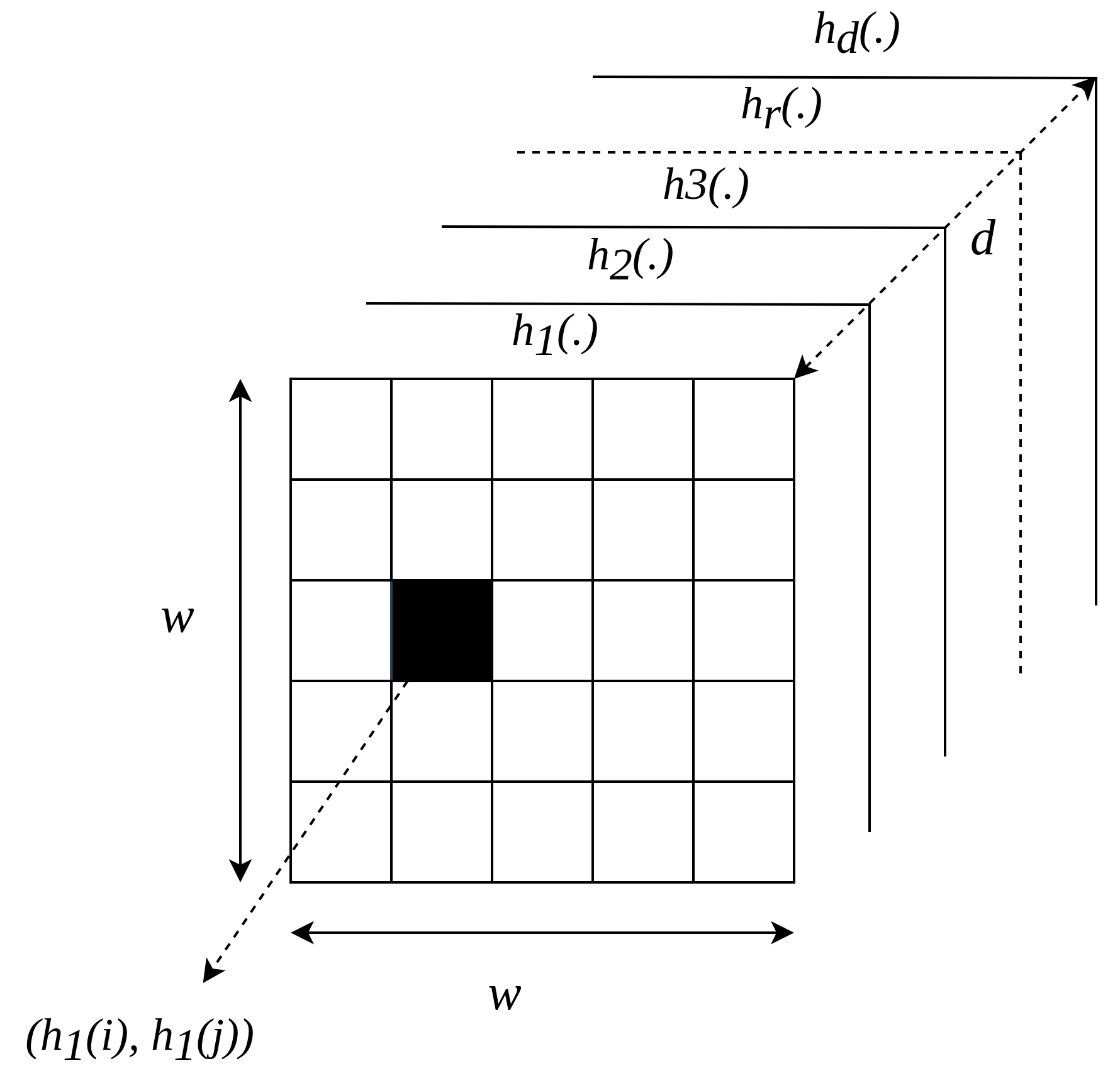}}
    \caption{TCM sketch\cite{khan_query-friendly_2016}}
    \label{fig:tcm}
\end{figure}

A disadvantage posed by all the approximate frequency count sketches like CountMin or gSketch is that they do not store the locality of the nodes. Therefore CountMin and gSketch cannot be used for conditional node queries or queries involving node connectivity. If these queries were to be run, the locality of the nodes has to be retained in the graph synopses. TCM\cite{tang_graph_2016} aims to solve this issue by storing the connectivity of the nodes in its data structure. TCM can summarize both node and edge information in constant time. Thus, it can answer a wide range of queries, unlike its predecessors. The structure of a TCM sketch is depicted in Fig.~\ref{fig:tcm}. TCM sketch could be considered as one of the pioneering works in summarizing data streams, which is directly related to our work presented in this paper.

\subsection{gMatrix}

The functionality of gMatrix\cite{khan_query-friendly_2016} is very similar to the TCM sketch. However, gMatrix considers several aspects which the TCM sketch does not address.

\begin{itemize}
    \item Reverse hashing queries through pairwise independent hash functions.
    \item Alternative options to extend sketch and space-saving synopses.
\end{itemize}

\section{Approach}

Through this research, we propose kMatrix, which is an improvement over the traditional gMatrix algorithm. The idea behind the kMatrix is to partition the 3-dimensional frequency matrix using a sample of the original graph steam as proposed in gSketch\cite{zhao_gsketch:_2011}. This idea has already been discussed in the TCM work to a certain degree. However, we explore this approach extensively with the gMatrix sketch, which can answer reachability queries in contrast to TCM. The significance of our approach is that kMatrix can answer all the queries that gMatrix is able to, while occupying the same amount of space as its counterpart gMatrix sketch. 

\subsection{kMatrix}

Let a stream be, \(G = \langle e_{1}, e_{2}, \ldots, e_{m} \rangle\). This can be mapped to a graph, G = (V, E) where \(V\) is the set of nodes and \(E\) is a set of edges as \(\{e_{1}, e_{2}, \ldots, e_{m}\}\). We can summarize this graph using a 3-dimensional matrix sketch\cite{khan_query-friendly_2016}. The straightforward choice would be to use a sketch similar to the one shown in Fig.~\ref{fig:tcm}. An edge, \((i, j) \in E\) will be hashed onto each layer of the sketch with has functions, \(h_{r}\), \(r \in \{1, \ldots, d\}\). The coordinate of the cell where the edge value is preserved will be \((h_{r}(i), h_{r}(j))\). Since the kMatrix aims to use the gMatrix sketch’s advantages over TCM, the hash functions should be pairwise independent of each other. 

However, by constructing a global sketch for the entire graph stream, some critical information about the structural properties of the underlying graph is dismissed. It is possible to improve the performance of a sketch by retaining some of these properties. Sketch partitioning\cite{zhao_gsketch:_2011} is one of the techniques that allow us to improve the sketch using the properties of the graph stream. In sketch partitioning, the global sketch is partitioned using a sample of the original graph stream such that it is possible to maintain a sufficient frequency uniformity within each partition. In this work, we use the sketch partitioning process discussed in gSketch to increase the accuracy of the queries further.

\begin{figure}[htbp]
    \centerline{\includegraphics[width=0.5\textwidth]{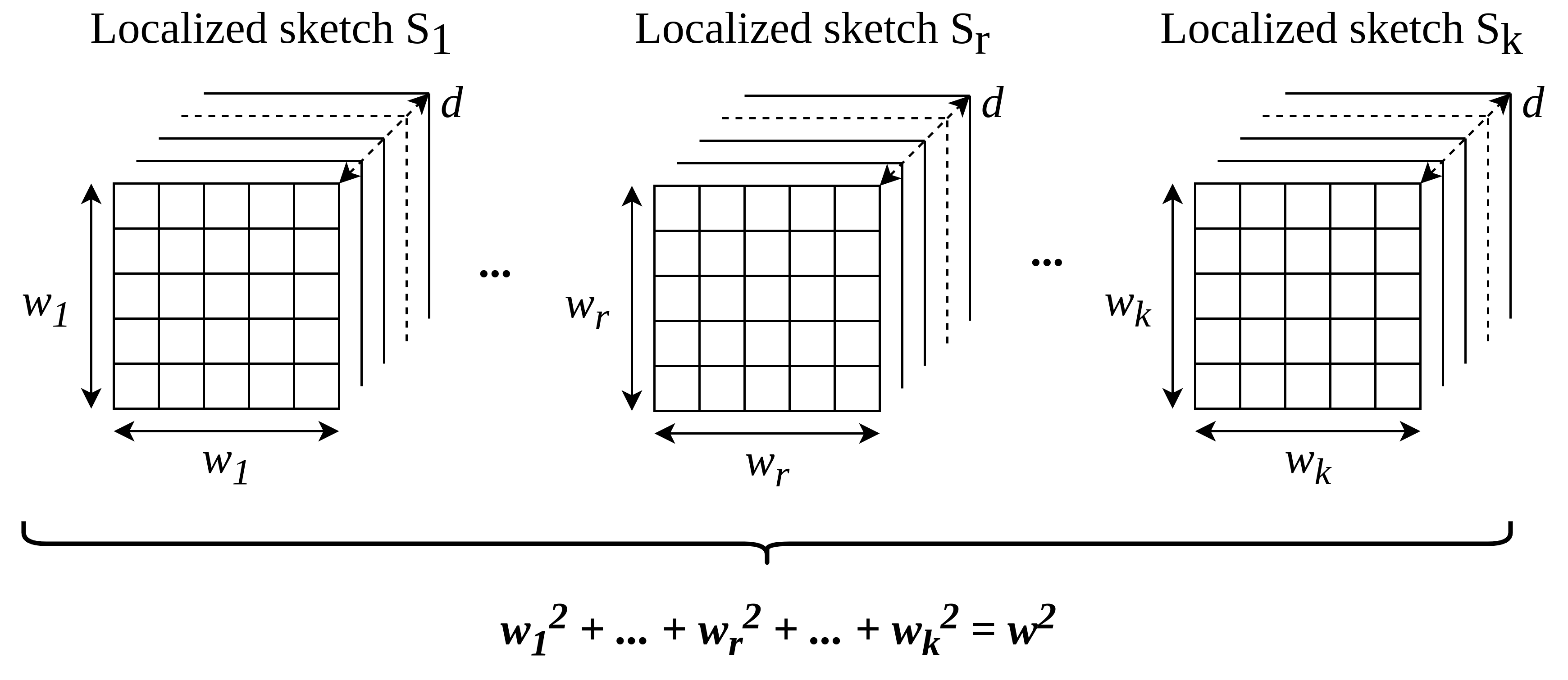}}
    \caption{kMatrix sketch}
    \label{fig:kmatrix}
\end{figure}

Fig.~\ref{fig:kmatrix} depicts the high-level view of kMatrix sketch after partitioning. Here, the sum of the memory occupied by all the localized sketches is equal to the memory allocated for the initial global sketch. The proceeding section will explain the partitioning algorithm in detail. 

\subsubsection*{Partitioning Algorithm}

Consider that the original sketch is partitioned into i sub-sketches. Let \(F(S_i)\) be the sum of the edge frequencies in the \(i\)th sketch. If \((m,n)\) is an edge in the \(i\)th sketch, let \(f(m,n)\) and \(\bar{f}(m,n)\) be its frequency and expected frequency respectively. Then,

\begin{equation}
    \bar{f}(m,n) = \frac{F(S_i) - f(m,n)}{w_i}
    \label{eq:1}
\end{equation}

Then the expected relative error of the edge \((m,n)\) is given by,

\begin{equation}
    \bar{e}(m,n) = \frac{\bar{f}(m,n)}{f(m,n)} = \frac{F(S_i)}{f(m,n) . w_i} - \frac{1}{w_i}
    \label{eq:2}
\end{equation}

The overall relative error \(E_i\) of the sketch can be expressed as the sum of the expected relative errors of all the edges.

\begin{equation}
    \sum_{(m,n) \in S_i}^{} \bar{e}(m,n)
    \label{eq:3}
\end{equation}

Let the average frequency of a vertex \(m\) be \(\tilde{f}_v(m)\) and the estimated out-degree be \(\tilde{d}(m)\). Then the average frequency of the vertex would be \( \tilde{f}_v(m) / \tilde{d}(m) \). Therefore the total estimated frequencies of the partitioned sketch \(S_i\) can be expressed as,

\begin{equation}
    \tilde{F}(S_i) = \sum_{m \in S_i \: ; \: m \in V}^{} \tilde{f}_v(m)
    \label{eq:4}
\end{equation}

According to the \eqref{eq:2}, \eqref{eq:3} and \eqref{eq:4},

\begin{equation}
    E_i = \sum_{m \in S_i}^{} \frac{\tilde{d}(m) . \tilde{F}(S_i)}{w_i . (\tilde{f}_v(m) / \tilde{d}(m))} - \sum_{m \in S_i}^{} \frac{\tilde{d}(m)}{w_i}
    \label{eq:5}
\end{equation}

\(\tilde{d}(m)\) in the numerator accounts for the fact that \(O(\tilde{d}(m))\) edges are coming out of the vertex m.

When a sketch of width \(w\) is partitioned into two sketches of widths \(w_1\) and \(w_2\), the total error can be expressed as \(E = E_1 + E_2\). Let \(w_1 = w_2\). Then,

\begin{multline}
    E = \sum_{m \in S_1}^{} \frac{\tilde{d}(m) . \tilde{F}(S_i)}{w_1 . (\tilde{f}_v(m) / \tilde{d}(m))} + \sum_{m \in S_2}^{} \frac{\tilde{d}(m) . \tilde{F}(S_i)}{w_2 . (\tilde{f}_v(m) / \tilde{d}(m))}\\ - \sum_{m \in S_1 \cup S_2}^{} \frac{\tilde{d}(m)}{w_1}
    \label{eq:6}
\end{multline}

The \eqref{eq:6} can be further simplified as,

\begin{equation}
    E' = E . w_1 + \sum_{m \in S_1 \cup S_2}^{} \tilde{d}(m)
\end{equation}

where the value of \(E'\) is,

\begin{equation}
    E' = \sum_{m \in S_1}^{} \frac{\tilde{d}(m) . \tilde{F}(S_i)}{\tilde{f}_v(m) / \tilde{d}(m)} + \sum_{m \in S_2}^{} \frac{\tilde{d}(m) . \tilde{F}(S_i)}{\tilde{f}_v(m) / \tilde{d}(m)}
    \label{eq:7}
\end{equation}

Thus it can be shown that the overall error can be minimized by choosing the smallest \(E'\) according to the \eqref{eq:7}. Therefore the underlying idea behind the partitioning algorithm is to choose a data sample of the original stream and then repeatedly partition the available space between the vertices in the sample according to the \eqref{eq:7}. After this partitioning phase, the streaming can begin and the edges that represented the vertices in the sample are put into their respective partitioned sketches.

A seperate data structure has to be used in order to track the vertices belonging to different localized partition. However the extra cost of storing this information is negligible when compared with the advantages obtained with the sketch partitioning.
\section{Evaluation}

\subsection{Experimental Setup}

\begin{figure}[htbp]
    \centerline{\includegraphics[width=0.5\textwidth]{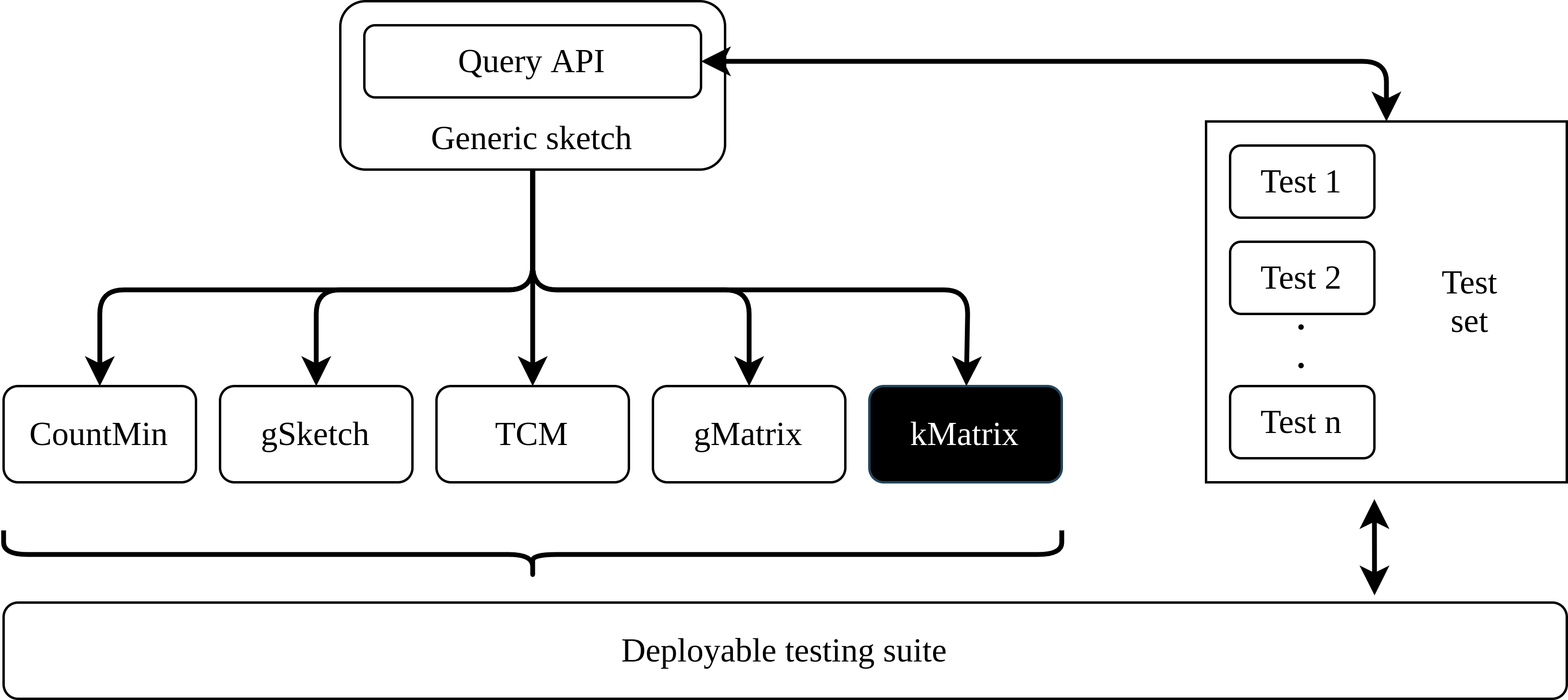}}
    \caption{High level architecture of the test suite}
    \label{fig:test_suite}
\end{figure}

The implementation mainly consists of two components; the test suite and the sketching algorithms. The entire code-base has been written in Python 3.8. The architecture of the benchmarking test suit is shown in Fig.~\ref{fig:test_suite}. It was designed in a modular way such that it permits easy addition of more test cases and graph sketches when necessary. 

All the tests were run on a 12-core Ryzen 3900 machine with a base clock of 3.1GHz and 32 GB RAM. However, only one core was utilized in running the tests.

A sample of 30,000 edges has been extracted from relevant datasets for initializing kMatrix at the beginning of each experiment. This sample stream has been obtained using reservoir sampling. 

\subsection{Datasets}

3 datasets were chosen to carry out the benchmarking process in this research. These were chosen to represent different application domains. 

\paragraph{unicorn-wget\cite{DVN/5H4TDI_2018}}
unicorn-wget is a dataset created from capturing the packet information of the network activity of a simulated network. This dataset was created at Harvard University. The dataset consists of 5 parts. From them, Hour-Long Wget Benign Dataset (Base Graph) which consist of 17,778 nodes and 2,779,726 edges was chosen for the experiment. We filtered 10\% of the edges using reservoir sampling for our experiments. 

\paragraph{email-EuAll\cite{leskovec_graph_2007}}
This data was extracted using email data from a large European research institution. The dataset consists of emails sent out in a period of 18 months. Each data item contains sender, receiver and the time of the origination of each email. The dataset consisted of 265,214 nodes and 420,045 edges\cite{noauthor_snap_nodate_email}. 

\paragraph{cit-HepPh\cite{leskovec_graphs_2005, gehrke_overview_2003}}
cit-HepPh citation graph is from the e-print arXiv regarding high energy physics phenomenology. It has 34,546 papers (nodes) and 421,578 citations (edges). We used the full dataset in our experiments. 

\subsection{Evaluation Metrics}
\label{section:design_evaluation_metrics}

\subsubsection{Average Relative Error (ARE)}
\label{section:metrics_are}

The average relative error is defined as,

\begin{equation}
    er(Q) =  \frac{\tilde{f}'(Q) - f(Q)}{f(Q)} = \frac{\tilde{f}'(Q)}{f(Q)} -1
\end{equation}

Given a set of m queries, $\{ Q_1 , ....., Q_m \}$, the average relative error is defined by taking the average of the relative error of all queries $Q_i$ for \(i \in [1,m]\).

\begin{equation}
    e(Q) =  \frac{\sum_{i=1}^{k} er(Q_i)}{m}
\end{equation}

\subsubsection{Number of Effective Queries (NEQ)}
\label{section:metrics_neq}

A query is said to be effective if the error, $\tilde{f}'(Q) - f(Q), < G_0$,  where $G_0$ is a predefined value. The number of effective queries is defined as,

\begin{equation}
    g(Q) =  \left | \{\,q\, |   \left |\tilde{f}'(q) - f(q)\right | \leq G_0, \,q \, \epsilon  \,Q\} \, \right|
\end{equation}

This can also be expressed as a percentage of effective queries (PEQ).

\begin{equation}
    g(Q) =  \frac{\left | \{\,q\, |   \left |\tilde{f}'(q) - f(q)\right | \leq G_0, \,q \, \epsilon  \,Q\} \, \right|}{|Q|}*100
\end{equation}

\subsection{Results}

This section will describe all the experiments conducted to measure the effectiveness of kMatrix against existing streaming graph sketching techniques. 

We have considered CountMin, gSketch, TCM, gMatrix and kMatrix sketches in our experiments. These sketches can be categorized into two groups depending on the type of queries they are able to answer.

\begin{enumerate}
    \item \emph{Type I} - The sketches which support only the edge frequency queries, i.e. CountMin and gSketch.
    \item \emph{Type II} - The sketches which support many graph queries in general, i.e. TCM, gMatrix and kMatrix
\end{enumerate}

Since \emph{Type I} sketches cannot answer anything other than edge frequency queries, we have only included \emph{Type II} sketches in our comparisons against kMatrix.

\subsection{Build-time}

\begin{figure}[htbp] 
    \centering
    \subfloat[unicorn-wget\label{fig:btt-a}]{\includegraphics[width=0.45\linewidth]{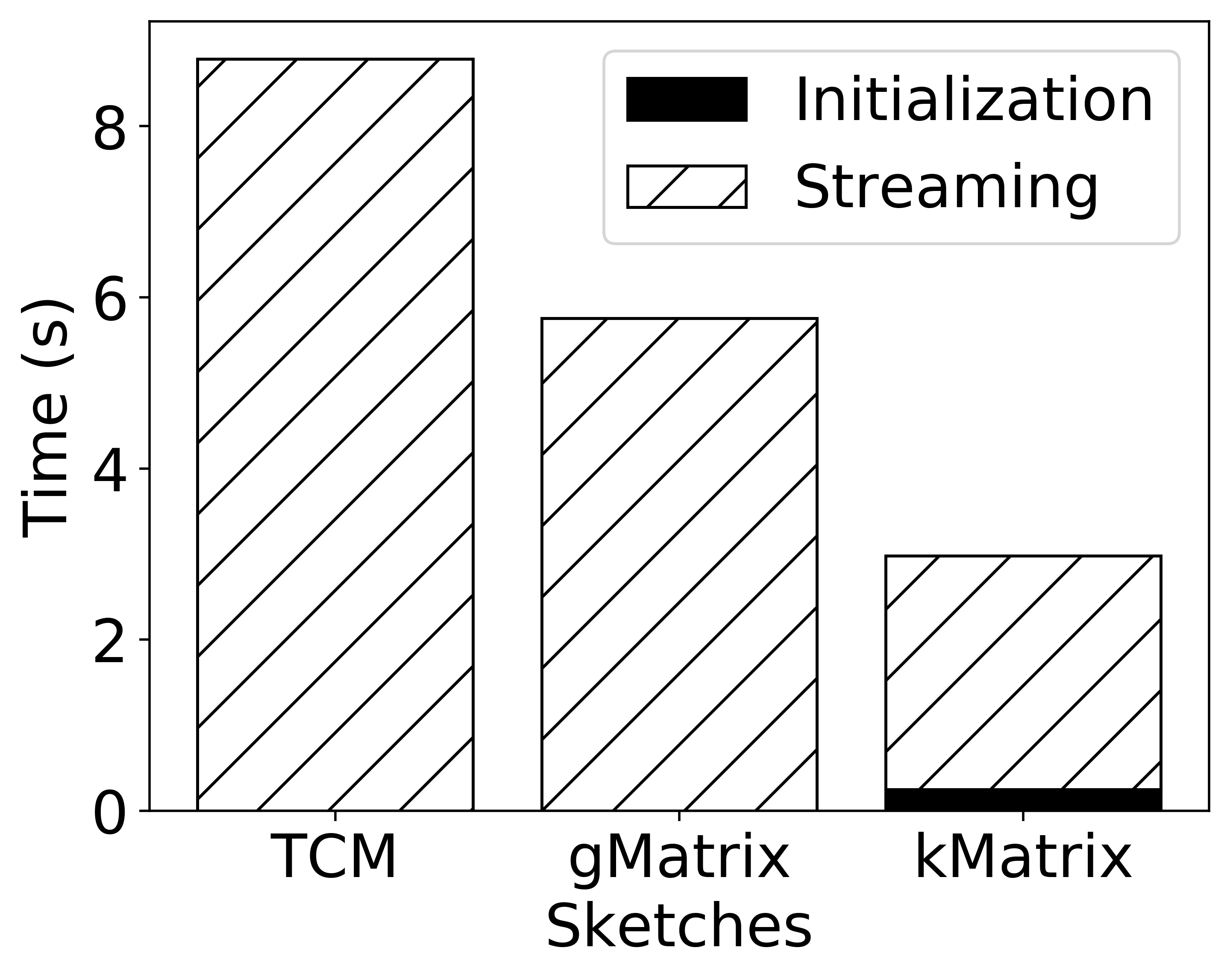}}
    \hfill
    \subfloat[email-EuAll\label{fig:btt-b}]{\includegraphics[width=0.45\linewidth]{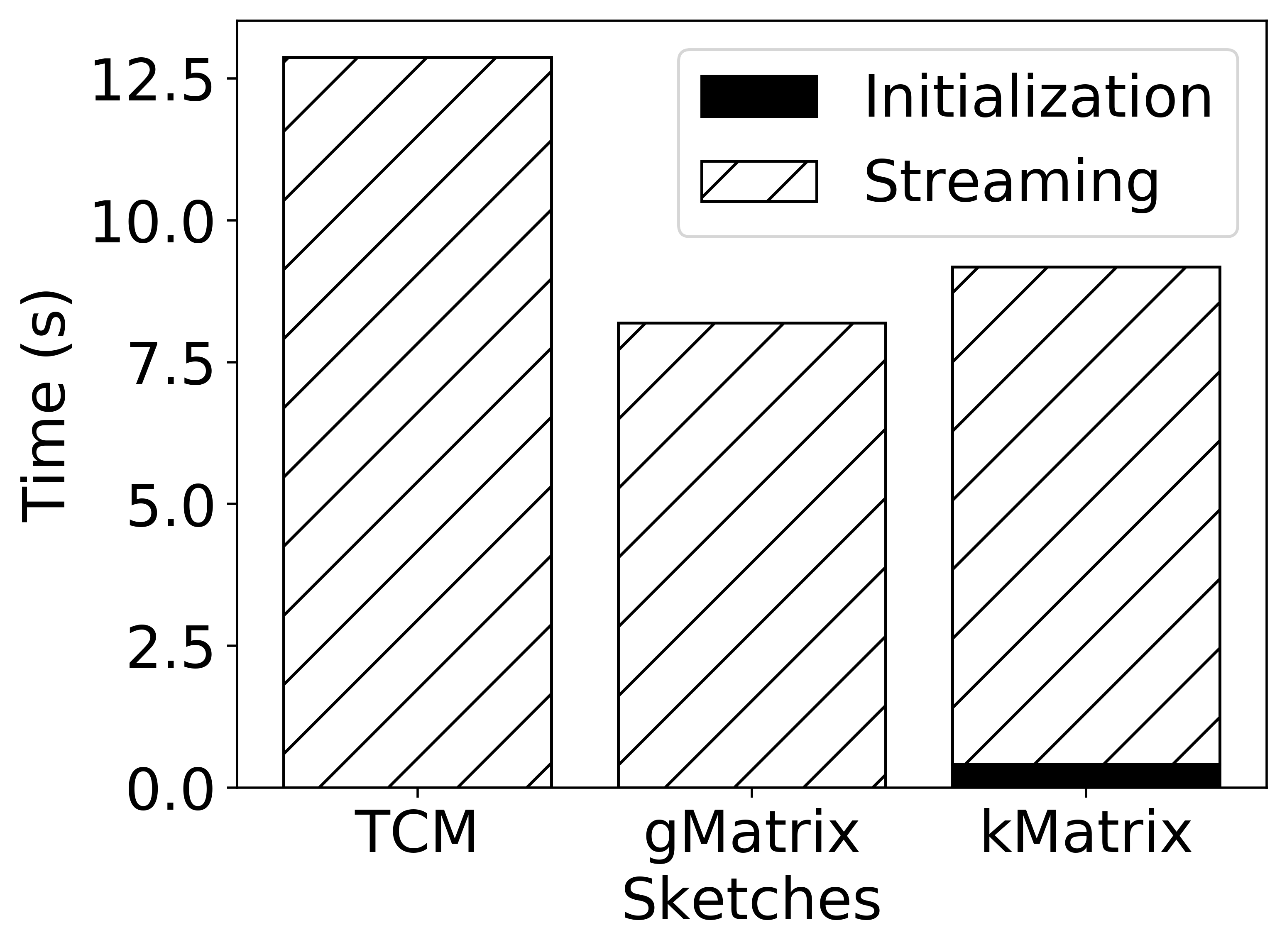}}
    \hfill
    \subfloat[cit-HepPh\label{fig:btt-c}]{\includegraphics[width=0.45\linewidth]{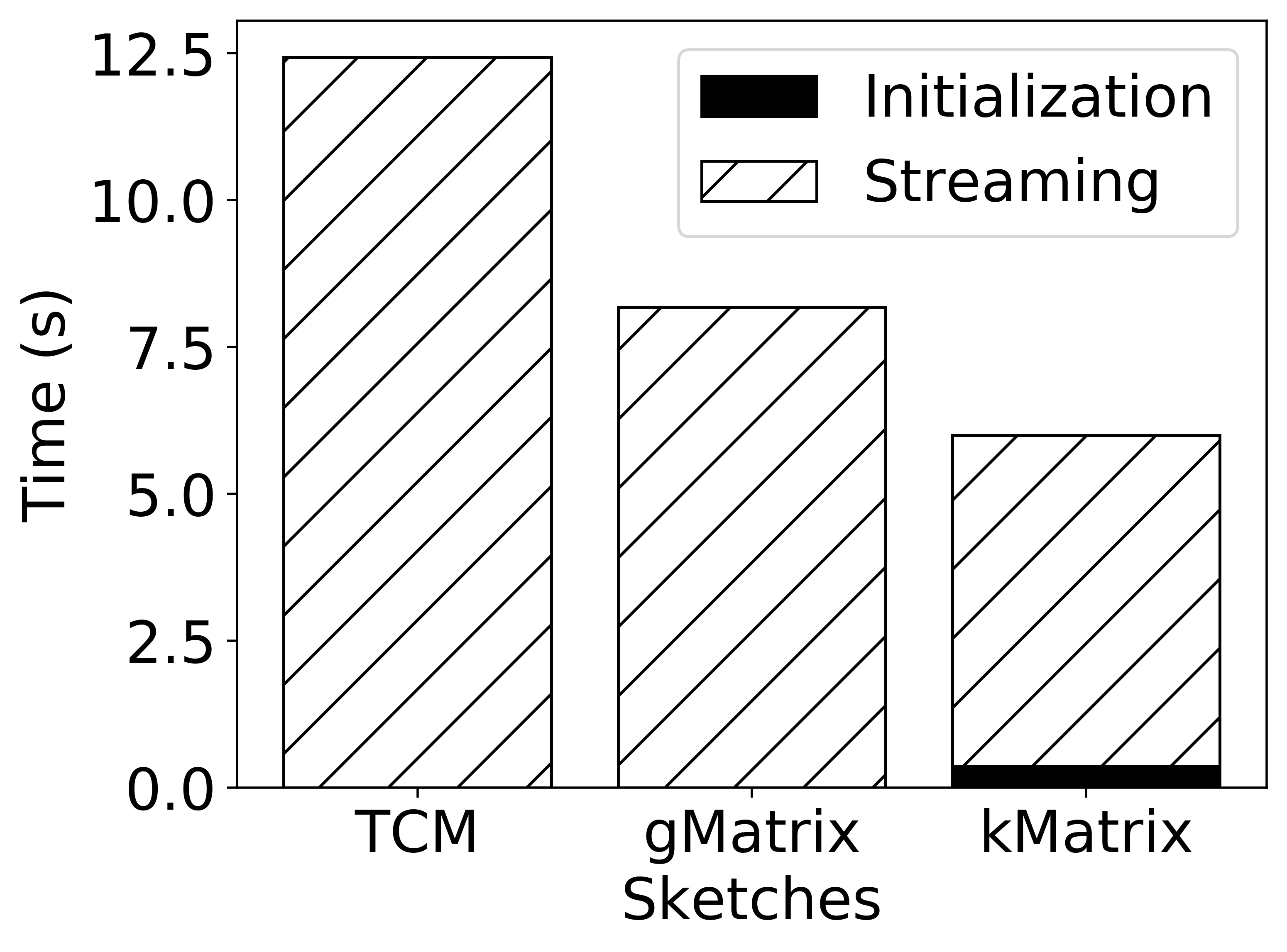}}
    \caption{Build-time}
    \label{fig:buildtime-test}
\end{figure}

Here we investigated the time to add the entire dataset to the sketch. The sketches were allocated a constant memory size of 1 MB, and the number of hash functions was set to \(d = 7\). The edges were streamed at the maximum throughput of each sketch. Therefore this experiment gives an idea about the average insertion rate of edges for each sketch. A minor drawback of kMatrix is that it takes some time for its initialization stage. However, this initialization time becomes negligible compared to the advantage that kMatrix receives over time due to its faster streaming rate. In both Fig.~\ref{fig:btt-a} and Fig.~\ref{fig:btt-c}, it has managed to outperform other sketching techniques by a significant margin. In Fig.~\ref{fig:btt-b}, kMatrix has shown comparable performance to gMatrix.

\subsection{Edge queries}

This experiment investigates how accurately the kMatrix can answer the edge queries after the summarization process. For this, we let our datastream get summarized into the sketch and then queried the frequency of different edges chosen at random. The experiment was repeated for each sketch for the sizes, 200 KB, 300 KB, 400 KB and 512 KB while keeping the number of hash functions at \(d = 7\). We have used average relative error and the number of effective queries as the evaluation matrices for this experiment.

\subsubsection{Average Relative Error}
\label{section:results_are}

\begin{figure}[htbp] 
    \centering
    \subfloat[unicorn-wget\label{fig:are-a}]{\includegraphics[width=0.45\linewidth]{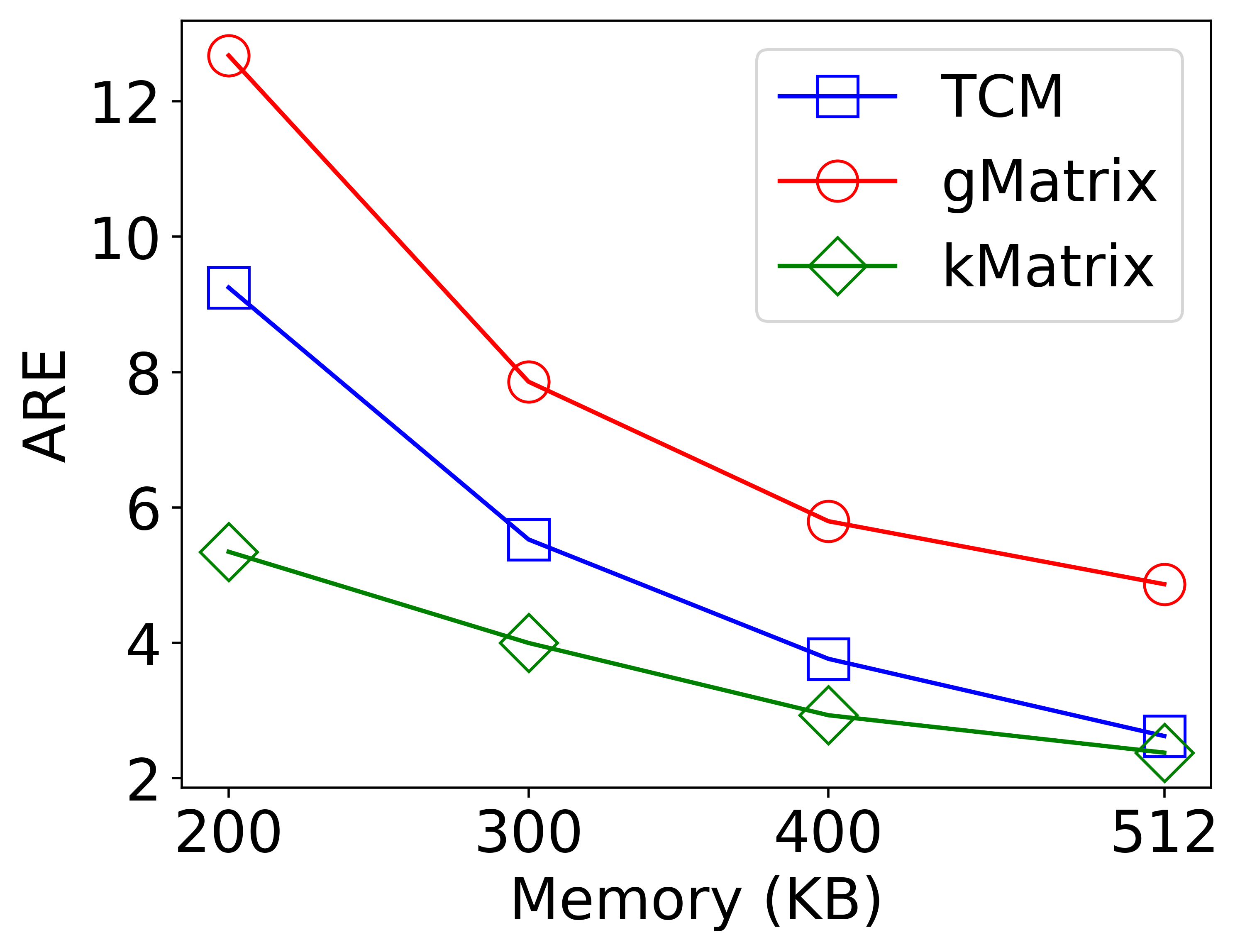}}
    \hfill
    \subfloat[email-EuAll\label{fig:are-b}]{\includegraphics[width=0.45\linewidth]{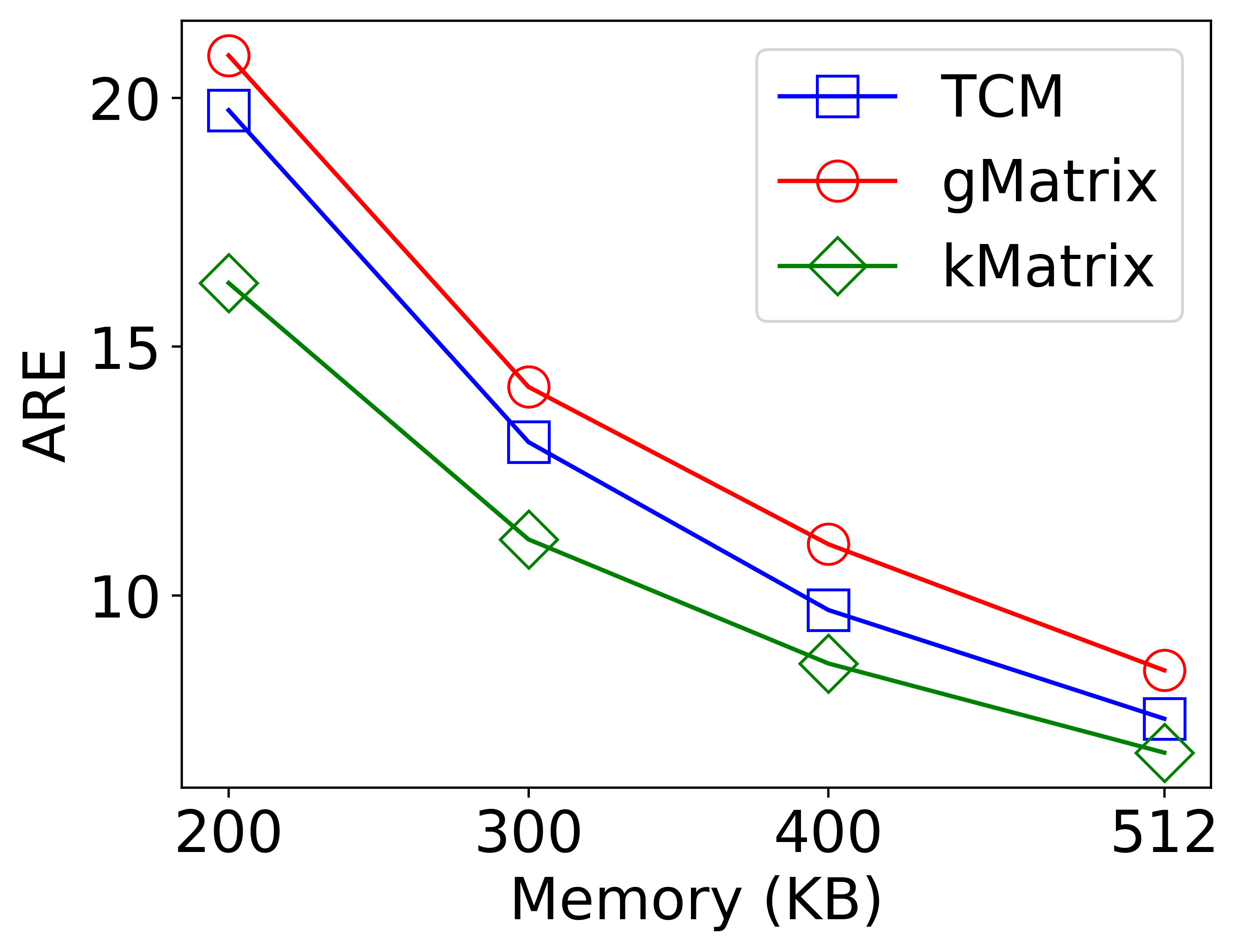}}
    \hfill
    \subfloat[cit-HepPh\label{fig:are-c}]{\includegraphics[width=0.45\linewidth]{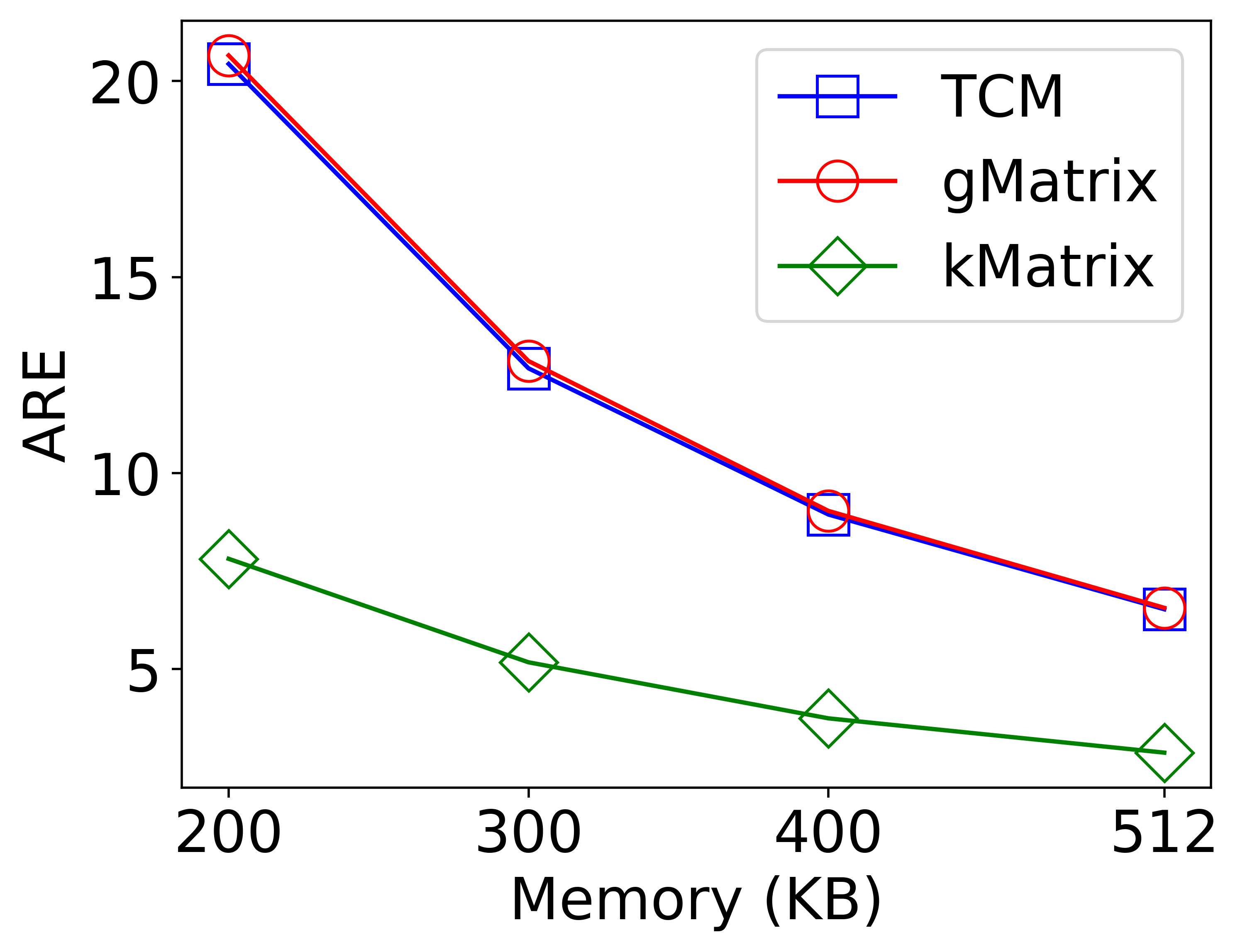}}
    \caption{Average relative error}
    \label{fig:edgq-queries-are-test}
\end{figure}

kMatrix showed significantly low ARE than all the other sketches for the three datasets we chose. The reason is that kMatrix can maintain frequency uniformity within each partition, making kMatrix relatively more immune to hash collisions than TCM and gMatrix. It is clear from the experimental evidence shown in Fig.~\ref{fig:edgq-queries-are-test} that kMatrix vastly outperforms the other state of the art sketching techniques. The superiority of our solution is more apparent when the allocated memory is low.

\subsubsection{Number of Effective Queries}

\begin{figure}[htbp] 
    \centering
    \subfloat[unicorn-wget\label{fig:neq-a}]{\includegraphics[width=0.45\linewidth]{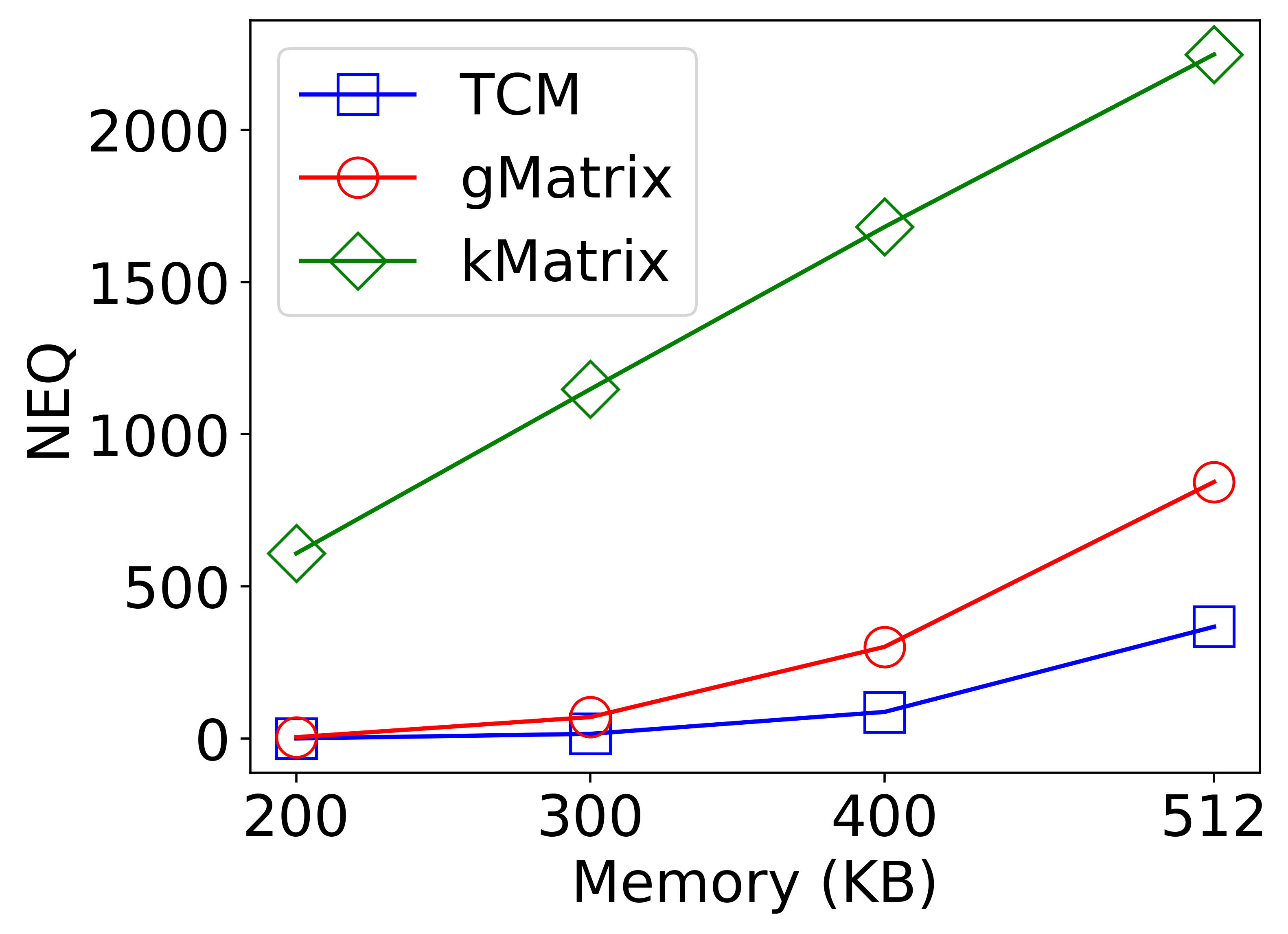}}
    \hfill
    \subfloat[email-EuAll\label{fig:neq-b}]{\includegraphics[width=0.45\linewidth]{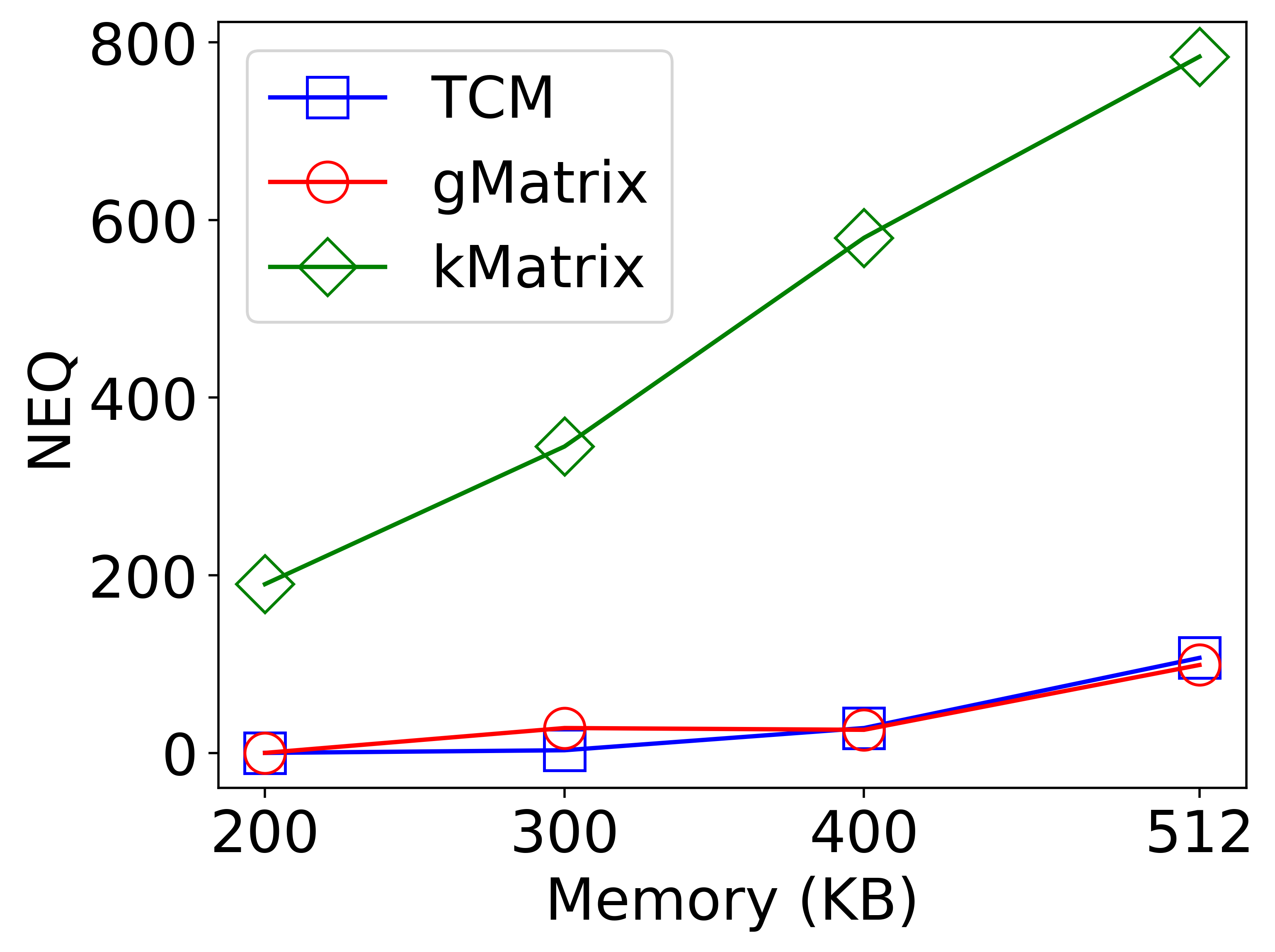}}
    \hfill
    \subfloat[cit-HepPh\label{fig:neq-c}]{\includegraphics[width=0.45\linewidth]{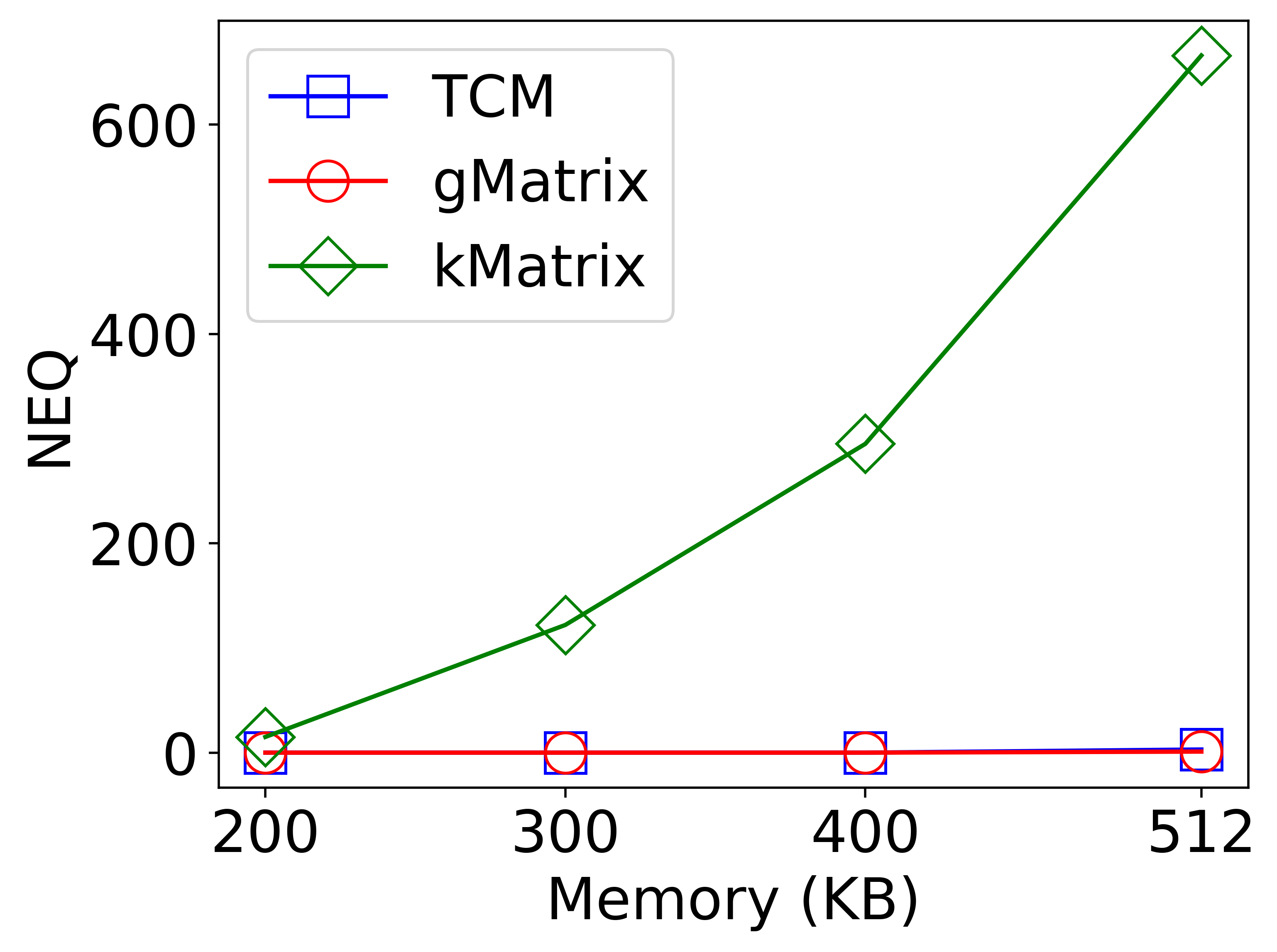}}
    \caption{Number of effective queries}
    \label{fig:edgq-queries-neq-test}
\end{figure}

The number of effective queries for each sketch was calculated by querying the sketches against 10,000 edges chosen through reservoir sampling from the original dataset. kMatrix has surpassed the accuracy of both TCM and gMatrix for all the scenarios that we have tested. The results for cit-HepPh in Fig.~\ref{fig:neq-c} shows that kMatrix has been able to effectively answer a significantly larger number of queries where the other sketches failed due to hash collisions.  
\section{Future work}

There are multiple aspects such as sliding windows and data partitioning across machines, that should be considered before the kMatrix sketch be used in a practical application. In addition, we have to test further the performance of kMatrix concerning other criteria such as heavy node/edge queries. The test suit’s functionalities can be extended and improved upon as a benchmarking tool for testing the graph summarization sketches. 
\section{Conclusion}

kMatrix is a new streaming graph summarization technique proposed through this research. It can answer queries with a significantly lower average relative error with the same amount of memory compared to the existing state-of-the-art sketching techniques, TCM and gMatrix. We have benchmarked kMatrix using three datasets in different application domains. We believe that the experimental results show the superiority of the proposed solution in comparison to the existing steaming graph summarization techniques. 

\bibliographystyle{ieeetr}
\bibliography{ms}

\end{document}